\documentclass[3p,twocolumn]{elsarticle}

\usepackage{graphicx,srcltx}

\def\fm3{\rm ~fm^{-3}}
\def\MeV{\rm ~MeV}	
\newcommand{\beq}{\begin{equation}}
\newcommand{\eeq}{\end{equation}}
\newcommand{\beqa}{\begin{eqnarray}}
\newcommand{\eeqa}{\end{eqnarray}}

\newcommand{\derc}[3]
{\left(\frac{\partial #1}{\partial #2}\right)
 \raisebox{-1em}{\ensuremath{#3}}}

\graphicspath{{eps/}}

\title{Low density symmetry energy effects in the neutron star crust properties}

\author{S.Kubis}
\address{H.Niewodnicza\'nski Institute of Nuclear Physics, Radzikowskiego 152, 
31-342 Krak\'ow, Poland}
\author{J.Por\c{e}bska}
\address{Jagiellonian University, Reymonta 4, 30-059 Krak\'ow, Poland}
\author{D.E. Alvarez-Castillo}
\address{H.Niewodnicza\'nski Institute of Nuclear Physics, Radzikowskiego 152, 
31-342 Krak\'ow, Poland}

\begin{document}

\begin{abstract} The form of the nuclear symmetry energy $E_s$ around saturation
point  density leads to a different crust-core transition point in the neutron
star and affect the crust properties. We show that the knowledge about $E_s$
close to the saturation point is not sufficient, because the very low density
behaviour is relevant. We also claim that crust properties are strongly
influenced by the  very high density behaviour of $E_s$, so in order to conclude
about the form of low density part of the symmetry energy one must isolate
properly  the high density part.

\end{abstract}


\maketitle

\section{Introduction}

One of the most intriguing quantity in the description of nuclear 
matter in neutron stars is the symmetry energy $E_s$, which is defined as
follows
\beq
E_{nuc}(n,x)=V(n) + E_s(n)\, \alpha^2 + {\cal O}(\alpha^4)
\label{Enuc}
\eeq
where  $E_{nuc}(n,\alpha)$ represents the energy of nucleonic matter 
per baryon as a function of baryon number density $n$ and the isospin asymmetry
$\alpha$, where $\alpha\!=\!(n_n\!-\!n_p)/n$ and $n_n,n_p$ are the neutron 
and proton densities.
At the  saturation point
density $n_0=0.16 \fm3$ the value of symmetry energy corresponds to the $a_4$ 
parameter in the Bethe-Weizs\"acker 
mass formula, and takes the value $E_s(n_0) = 30\pm 1\MeV$. Isoscalar part  of
interactions is represented by the isoscalar potential $V(n)$ which is mainly
responsible for the stiffness of the Equation of State (EoS).

Density dependence of 
$E_s$ is however highly uncertain both below and above saturation point $n_0$. 
This dependence is one of the goals of the experimental investigations
carried on the radioactive beam colliders \cite{Baran:2004ih,Li:2008gp}. This
 kind of facilities
allow for research of nuclear matter with large isospin asymmetry.
The analysis shown in  \cite{Chen:2004si,Chen:2005ti,Chen:2007qb}
put some constraints on the slope and curvature of $E_s(n)$ around
$n_0$ however we are still far from the final conclusion about the global shape of
the symmetry energy.

The role played by the $E_s$ in the context of various neutron star observables
was emphasized in \cite{Lattimer:2000nx}   and more detailed analysis was made
in \cite{Steiner:2004fi}. One of the first approach to the crust-core
transition  was presented in  \cite{Baym:1971ax} where the stability
considerations were performed. This kind of analysis with an improved nuclear
model was later used  in  \cite{Pethick:1994ge}. In this work  authors 
suggested the different critical density in different nuclear models comes from
their discrepancy in the neutron matter description, e.g. in the symmetry energy
form. The direct connection of $E_s$ to the crust-core transition point
was shown in \cite{Kubis:2006kb} and possible phase separation in the inner
 core was analysed in \cite{Kubis:2007zz}. 
In this work we would like to go along this line to emphasize that the very low
density behaviour of the symmetry energy is essential for the crust-core
transition point. It is especially interesting taking into account the recent 
experimental results \cite{Kowalski:2006ju,Natowitz:2010ti} which show the
symmetry energy still takes large values at densities very much below $n_0$.
This result is in contrast to common conviction coming from various theoretical
approaches that states the $E_s$ goes almost linearly to zero for low densities.
So, the consequences of the symmetry energy with such unusual feature 
seems to be worth to be seen.

\section{The crust-core transition point}

In order to estimate the crust-core transition point  the bulk instability
conditions were applied. It is based on the vanishing compressibility which
signals the one-phase  system is unstable against the density fluctuations and 
must split into two phases. It was shown in \cite{Kubis:2006kb} that 
the compressibility
under constant charge chemical potential - $K_\mu$ is the proper quantity 
in case of matter under beta equilibrium. It is defined as  follows
\beq
K_\mu = \derc{P}{n}{\mu}
\eeq
where $P$ is the total pressure (nucleons + leptons).
When nuclear  matter contribution to the total
energy is described by Eq. (\ref{Enuc}) then the stability condition takes the
form 
\beqa
K_\mu&\!=\!&
n^2\left(E_s'' \alpha^2+V''\right) + 
2\; n\!\left(E_s' \alpha^2+V'\right)  \nonumber \\ 
&&-\frac{2 \alpha^2 E_s'^2 n^2}{E_s} \ge 0
\label{stab}
\eeqa
where the role played by the symmetry energy $E_s$ is apparent. For densities
typical  for the neutron star core
the compressibility $K_\mu$ is positive but decreases
as density decreases. For some density $n_c$, located below $n_0$,
the compressibility
vanishes. It means the charge fluctuations are not stable - below $n_c$
the matter cannot exist in one phase and must separate into 
two phases which may form a crystal lattice.
However, the answer, where is situated the inner edge of the crust, is much more
complicated.

\begin{figure}[h]
\includegraphics[clip,width=.95\columnwidth]{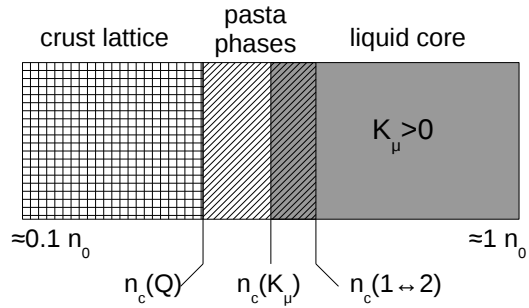}
\caption{Schematic picture of the different estimation for the crust-core
transition in the neutron star. The gray region corresponds to $K_
\mu<0$, the uncertain hatched range indicates region where pasta phases 
may occur.}
\label{c-c}
\end{figure}

There is no unique
method to determine at what density the solid crust starts to form. Here we
want to discuss this issue.
The critical density for vanishing compressibility  $n_c(K_\mu)$
 represents the absolute limit for the homogeneous and neutral
system. It means the phase splitting must occur before reaching the $n_c$.
Indeed, the energy for the two-phase system becomes smaller than 
for the one-phase system before the density approaches the critical value 
$n_c(K_\mu)$  \cite{Douchin:2000kx,Pethick:1994ge}.
 The boundary for two phase coexistence, 
$n_c(1\leftrightarrow  2)$ is then located slightly above $n_c(K_\mu)$ as 
it is shown in Fig.\ref{c-c}.
Both $n_c(K_\mu)$ and $n_c(1\leftrightarrow  2)$ are quantities derived in  the
bulk approximation, i.e. without the finite size effects included. The 
finite size structures require inclusion of 
the additional forms of energy like the Coulomb and  surface  energy.
Competition between them, leads to structures with various dimensionality (rods,
plates) usually called pasta phases \cite{Lorenz:1992zz}. The presence of these
exotic structures is however model dependent, moreover mechanical properties of
rods and plates resemble rather liquid crystals than solid state
\cite{potekhin}. The shear tensions in all spatial directions can be supported
only by a 3-dimensional lattice of 0-dimensional structures, 
e.g. the lattice of almost point-like nuclei immersed in the less dense
medium, e.i. the gas of electrons  and dripped neutrons.
So, for the most reliable position of the crust edge one should
take the point where the nuclei start to deform into very long structures or,
keeping their spherical form, dissolve in the surrounding medium. 

The finite size effects may also be included in the stability consideration of
one-phase system as was done in \cite{Pethick:1994ge}. The most important
ingredient in this analysis appeared to be the Coulomb energy.
Being repulsive, the Coulomb force stabilizes fluctuations of charge and 
makes matter more resistive to formation of charged clusters. The corresponding
critical density $n_c(Q)$ is then moved to lower densities,
 always below $n_c(K_\mu)$ (see Fig.\ref{c-c}).
For some kind of interactions (e.g. Skyrme-like in \cite{Douchin:2000kx})
the  $n_c(Q)$ almost coincides with the point where nuclei disappear, so
it may be treated as the very likely estimation of the crust-core transition.

In this work we stay with the simplest condition, $n_c(K_\mu)$ what means we 
get upper bound on the crust mass and thickness. The difference between 
$n_c(K_\mu)$ and $n_c(Q)$ is not so large, usually smaller than 10\%.
The comparison between these quantities 
was already done in \cite{Pethick:1994ge} and for modern
nuclear models in \cite{Oyamatsu:2006vd,Xu:2008vz}. 
The condition (\ref{stab}) corresponds to the cross-point of the border of
spinodal region at zero temperature with beta equilibrium condition in stellar
matter as it was considered in \cite{Ducoin:2008xs}.

\section{Nuclear models}

\begin{figure}[t]
\includegraphics[clip,width=.99\columnwidth]{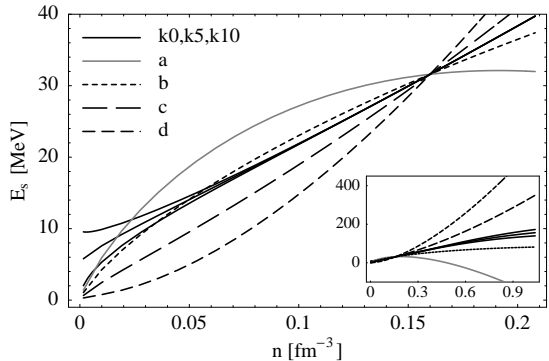}
\caption{The shape of symmetry energy for different models in the low and high 
density region (the inset).}
 \label{eos-fig}
\end{figure}
Different models of nuclear interactions, while being consistent at saturation
 point
$n_0 = 0.16 \fm3$, leads to large discrepancies
for EoS at higher densities. In order to estimate the role played by $E_s$
we used a model which have the same isoscalar part of interactions $V(n)$ 
and different form of the isovector part $E_s(n)$
The $V(n)$ taken from Prakash et.al. \cite{pal} correspond the
compressibility at saturation point $K_0=240~\rm MeV$.

Various shapes of symmetry energy, where taken from   MDI model
\cite{Chen:2004si} and named $a,b,c,d$ as in \cite{Kubis:2006kb}. The symmetry energy in MDI family
presents large discrepancies both in the low and in the high density region.
Also the slope $E_s'(n_0)$ and curvature $E_s''(n_0)$ at the saturation point
take different values. The only  common property is the  value of the  symmetry
energy at $n_0$. Especially, the large differences at high densities lead to a
differences in the global stellar parameters like its radius and compactness.
Those global parameters indirectly affect the crust thickness and its 
contribution to the total star mass what obscure the picture. 
In order to isolate the low density properties of $E_s$ from the rest 
we use a specific parametrization which presents the same shape of $E_s$ 
 at saturation point and very tiny differences at high
density.
It has the following form
\beq
\!\!E_s =  \frac{\alpha u^2+ \beta u + \gamma
\sqrt{u}+E_s(0)}{u^2+ 1}.
\eeq
where $u=n/n_\infty$.
The $n_\infty  $ control the behavior at very high density and was taken here 
$n_\infty = 1 \fm3$ to get intermediate values of $E_s$ between extremal
cases ($a$ and $d$) in the MDI family.
The remaining parameters $\alpha, \beta, \gamma$ were derived to ensure  
the same slope $L=80$~MeV and the stiffness $K_{asy} = -450$~MeV at $n_0$ 
while  $E_s$ at $n \rightarrow 0$  approaches three different 
values 0, 5 and 10 MeV. The shape of symmetry energy for all models
are shown in Fig.\ref{eos-fig}.

Above models of nuclear interaction cannot be use to obtain EoS for very 
low densities, in the crust region. In order to obtain 
the EoS int the full range, up to the edge of the star,  they have to be 
completed with the EoS for the crust itself.
Here we used  well established results  from 
\cite{hp} and \cite{dh}. 

\begin{table}
\caption{The $E_s(n)$  parametrization for $k$ models: 0,5,10. The last columns
presents the critical density.}
\label{param}
\begin{tabular}{ccccc}
$\alpha$ & $\beta$ & $\gamma $ & $E_s(0)$~[MeV] & $n_{c}$ \\
247.08 & 40.93& 48.836& 0 & 0.080 \\
181.98& 103.43& 15.502& 5 & 0.073 \\
116.87& 165.93& -17.831& 10 & 0.065
\end{tabular}
\end{table}

\section{Results}

\begin{figure}[b]
\includegraphics[clip,width=.85\columnwidth]{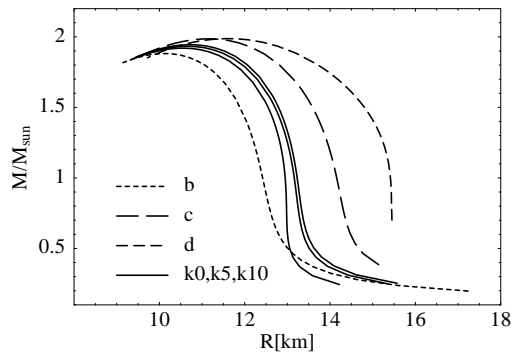}
\caption{The total star mass versus star radius.}
\label{R-M.eps}
\end{figure}
\begin{figure}[t]
\includegraphics[clip,width=.85\columnwidth]{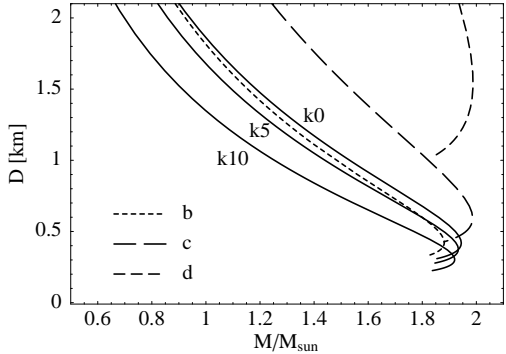}
\caption{The star crust thickness versus the total star mass.}
\label{Dcrust-M.eps}
\end{figure}
\begin{figure}[t]
\includegraphics[clip,width=.85\columnwidth]{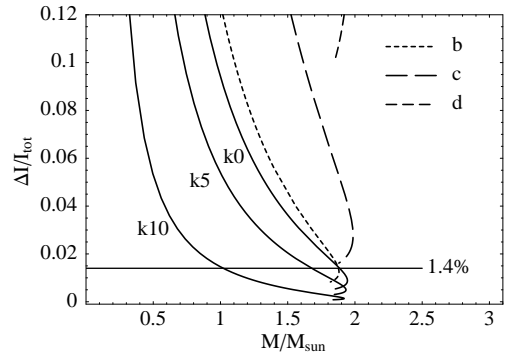}
\caption{The crustal fraction moment of inertia  versus the total star mass. The
lower bound, 1.4\%, for $I_{crust}/I$ coming from observations is indicated by the
horizontal line.}
\label{Icrust-M.eps}
\end{figure}

In this  section the results are presented. In the Fig.\ref{R-M.eps} the
mass-radius relation is shown. One may observe huge differences in  the  MDI
family. These differences are caused only by the symmetry  energy shape because 
the isoscalar part is the same for all EoSs. The symmetry energies in the MDI
family diverge  enormously at higher  densities. There are no results for the
$a$ model because its $E_s$ takes so large negative values at high density that
makes  the pressure bounded from above and leads to  unstable EoS. The rest,
$b$-$d$ models,  present large  differences in the compactness of neutron star
giving difference in the surface gravitation. Hence any crustal parameters like
thickness, mass etc. are changed not only by the form of $E_s$ at the crust
region but also by its values in deep  core region. For the $k$ family $E_s$ is
almost the same at high density and  the M-R relations for $k0$-$k10$ models are
similar to each other. The differences between them slightly increase with the
dropping mass what is natural as the crust occupies more an more volume of the
star and differences in the EoS at low density are  more pronounced. The effect
of the high density $E_s$ form  is apparent when one  compare critical densities
for different models. In the MDI family, for $b,c,d$ one get $n_c = 0.092,
0.095, 0.160$ (see \cite{Kubis:2006kb}) whereas for $k0,k5,k10$:  $n_c = 0.080,
0.073, 0.065 $ (see Tab.\ref{param}).  The  difference in critical density for
$b$ and $c$ is less then in the $k$ family but the difference in the crust
thickness (see Fig.\ref{Dcrust-M.eps}) is much larger  than
one could expect from the $n_c$ position. Surely it comes from the different
compactness of the star as a whole. In the $c$ case the star is more compact and
its crust is more squeezed by the gravity. The effect is even better seen for the
 crustal fraction of the moment of inertia, Fig.\ref{Icrust-M.eps}. 
As one can note for the typical $1.5 M_\odot$ star the relative 
 differences among $b,c,d$
models reach hundreds of percents  whereas for the $k$ family  are not so large. 
At the end of the discussion we focus on the  $k$ models themselves. The symmetry
energy for this family was constructed to ensure the same behaviour at high
density and keep the same slope and curvature at the saturation point in
agreement with recent results from experimental results. The only difference was
the asymptotic value of the $E_s$ at vanishing density $n \rightarrow 0$. We
have probed values 0, 5 and 10 MeV. Considering  the shapes of $E_s$in the
Fig.\ref{eos-fig}   one may conclude that the $E_s(0)$ is very essential 
quantity for the crust properties. The symmetry energies for the $k$ family are
almost overlapping in the $n_0$ region and below and their discrepancies seems
to be negligible in comparison to $b$-$d$ lines, however  they effect in quite large
differences for the  crust thickness and moment of inertia. We also observe that
higher $E_s(0)$ systematically makes the crust thinner and its contribution to 
total moment of inertia takes very low values.

 In order to refer our results to 
observations we indicated the upper bound on the crustal moment of inertia 
for the Vela pulsar coming from analysis  performed by Link in \cite{Link:1999ca}.
The line for $k10$ model  goes below $1.4 \%$ in very wide range of stellar
masses. We do not know the Vela pulsar mass, so in that sense the model cannot
be verified, but if we trust experimental results that $E_s(0) \approx 10$~MeV
it would mean the Vela pulsar mass should have its mass not greater than $1
M_\odot$. 

\section{Summary}

We perform the analysis on how the very low symmetry energy
behavior effects the basic neutron star crust parameters. Although the crust
encompasses the densities below $n_0$, also the high density  behavior
 of the symmetry energy influences its properties.
In order to be able to conclude about the low density part we constructed a special  
family of symmetry energies. They presented the same shape  at saturation
point and above but differ when density goes to zero. It was shown that unusual
property like the  non-vanishing  symmetry energy at zero density leads to 
a very thin crust with very  small contribution to the total moment of inertia.

\end{document}